\documentclass[12pt]{article}

\def\pt{$p_T$}
\def\ph{$\phi$}
\def\om{$\Omega$}
\def\la{$\Lambda$}
\def\rec{recombination }
\def\op{$\Omega/\phi$}
\def\dis{distribution}

\begin{document}

\begin{center} {\Large {\bf Production of strange particles at intermediate \pt\ \\  at RHIC }}
\vskip .75cm
       {\bf   Rudolph C. Hwa$^1$ and  C.\ B.\ Yang$^{1,2}$}
\vskip.5cm

       {$^1$Institute of Theoretical Science and Department of
Physics\\ University of Oregon, Eugene, OR 97403-5203, USA\\
\bigskip
$^2$Institute of Particle Physics, Hua-Zhong Normal University,
Wuhan 430079, P.\ R.\ China}
\end{center}
\vskip.5cm
\begin{abstract} 
  The recombination model is applied to the production of $K$, \ph, \la\ and \om\ at all \pt\ in central Au+Au collisions. The thermal-shower component of the recombination is found to be important for $K$ and \la, but only in a minor way for \ph\ and \om\ in the intermediate to high \pt\ region. The normalization and inverse slope of the thermal partons in the strange sector are determined by fitting the low-\pt\ data.  At  higher \pt\  the data of $K$, \la, \ph\ and \om\  in the log scale are all  well reproduced in our study that extends the thermal contribution and includes the shower contribution. The calculated result on the $\Lambda/K$ ratio rises to a maximum of around 2 at $p_T\approx 4$ GeV/c, arching over the data in linear scale.  The production of \ph\ and \om\ are shown to arise mainly from the recombination of thermal partons, thus exhibiting exponential \pt\ dependences in agreement with the data. Their ratio, $R_{\Omega/\phi}$, rises linearly to $p_T\approx 4$ GeV/c and develops a maximum at $p_T\approx 5.5$ GeV/c. It is argued that the \pt\ spectra of \ph\ and \om\ reveal directly the partonic nature of the thermal source that characterizes quark-gluon plasma. Comments are made on the \om\ puzzle due to the simultaneous observation of both the exponential behavior of the \om\ spectrum in \pt\ and the existence of low-\pt\ particles associated with \om\ as trigger.

\end{abstract}

 \section{Introduction}
 The production of strange particles has always been a subject of great interest in heavy-ion collisions because of their relevance to possible signatures of deconfinement and flavor equilibration \cite{1,2}. Strangeness enhancement that has been observed at various colliding energies is a phenomenon associated with soft particles in the bulk matter \cite{3,4}. At high transverse momentum ($p_T$), on the other hand, the production of jets does not favor strange particles, whose fragmentation functions are suppressed compared to those for non-strange particles. At intermediate \pt\ range between the two extremes the \pt\ distribution depends sensitively on both the strangeness content and the production mechanism. It has been shown that in that \pt\ range the spectra of $\pi$ and $p$ can be well described by parton recombination \cite{5}. In this paper we study the production of the openly strange particles $K$, \la\ and \om, as well as the hidden strange hadron \ph. The aim is to get a broad view of strangeness production at intermediate and high \pt\ in a unified framework.   
 
 The four particles emphasize four different aspects of hadronization: $K$ and \la\ involve non-strange quarks, whereas \ph\ and \om\ do not; $K$ and \ph\ are mesons, whereas \la\ and \om\ are baryons, although \ph\ and \la\ are comparable in masses. To get all four spectra correctly at all \pt\ would require a high degree of coordination in the theoretical description. We have already a fairly good description of the production of hadrons in the non-strange sector \cite{5}; however, for the strange sector we must deal with a new set of problems. Gluon conversion feeds the $s$ quarks in the thermal medium and leads to strangeness enhancement \cite{6}.  On the other  hand, the production of $s$ quarks at high \pt\ in jets is suppressed. The  interplay between the enhanced thermal $s$ partons and the suppressed shower $s$ partons results in very unusual outcome. Furthermore, how the supply of $s$ quarks is partitioned into various channels of hadrons containing open and hidden strangeness is an aspect of the recombination process that was not encountered in the non-strange sector, but was studied algebraically by quark counting in Refs.\ \cite{7,8} without gluon conversion. Our consideration here treats the problem in a much wider scope where the \pt\ \dis s of all four strange particles are investigated. 
    
We shall use the basic recombination formula to compute the \pt\ \dis s of all four particles. The thermal $s$ quark \dis\ is determined by fitting the low-\pt\
 data of $K$ production. The shower parton \dis s (SPD) are known \cite{9}. 
 Thus one can proceed to the calculation of the \la\ spectrum essentially without adjustable parameters. In the cases of \ph\ and \om\   the competition for multiple strange quarks to form such high mass states would lead to suppression. Moreover, since the $s$ quarks from the showers are suppressed compared to non-strange shower partons, \ph\ and \om\ produced at large \pt\ do not benefit from hard scattering as much as $K$ and \la\ do, for which light shower partons can contribute. 
The consequence is that the thermal partons play a dominant role --- to the extent that the production of \ph\ and \om\ provides a window to look into the thermal nature of the quark-gluon plasma before hadronization.  
 All these problems will be examined quantitatively in this paper.

 \section{Formulation of the Problem}
 
 We shall assume that all hadrons produced in heavy-ion collisions are formed by \rec of quarks and/or antiquarks, the original formulation of which is given in \cite{10,11} for $pp$ collisions. In recent years the recombination model has been extensively studied by many groups \cite{12,13,14,5} with great success in reproducing the \pt\ spectra in the intermediate \pt\ region of Au+Au collisions. In our 1D description of the \rec process
  the invariant inclusive distribution of a produced meson with momentum $p$  is
 \begin{eqnarray} 
 p^0{dN_M  \over  dp} = \int {dp_1 \over  p_1}{dp_2
\over p_2}F_{q\bar{q}'} (p_1, p_2) R_M(p_1, p_2, p) ,
\label{1}
\end{eqnarray} 
 and for a produced baryon
\begin{eqnarray}
p^0{dN_B\over dp}=\int {dp_1\over p_1}{dp_2\over p_2}{dp_3\over p_3}\,
F_{qq'q''}(p_1,p_2,p_3)\,R_B(p_1,p_2,p_3,p) .
\label{2}
\end{eqnarray}
 The properties of the medium created by the collisions are imbedded in the joint quark distributions $F_{q\bar q'}$ and $F_{qq'q''}$. The recombination functions (RFs) $R_M$ and $R_B$ depend on the hadron structure of the particle produced, but for a high mass hadron the normalization can in addition be influenced by the competition among possible channels. In the valon model description of hadron structure \cite{11,15}, the RFs are
 \begin{eqnarray} 
 R_M(p_1, p_2, p) =g_{M}\,y_1y_2\,  G_M (y_1, y_2 ) ,  \label{3}\\
 R_B(p_1,p_2,p_3,p)=g_{B}\,y_1y_2y_3\,G_B(y_1,y_2,y_3) , \label{4}
\end{eqnarray} 
 where $y_i=p_i/p$, and $g_M$ and $g_B$ are statistical factors. $G_M$ and $G_B$ are the non-invariant probability densities of finding the valons with momentum fractions $y_i$ in a meson and a baryon, respectively.
 	Equations (\ref{1}) and (\ref{2}) apply for the produced hadron having any momentum $\vec p$. We consider $\vec p$ only in the transverse plane, and write $p_T$ as $p$ so that $dN/p_Tdp_T$ becomes $(pp^0)^{-1}$ times the right-hand sides of Eqs.\ (\ref{1}) and (\ref{2}). 
	
	The wave function $G(y_i)$ in terms of the momentum fractions of the valons cannot be probed for $K, \phi, \Lambda$, and $\Omega$, as one can for proton. However, we can make reasonable estimates. For  $K$ and \la, there is a mixture of strange and non-strange constituents with different masses, so the corresponding momentum fractions are different. The wave function of $K$ has been considered before \cite{16}; it has the form
\begin{eqnarray}
G_K(y_1,y_2)={1\over B(a+1,b+1)}\ y_1^a y_2^b\,\delta(y_1+y_2-1) .   \label{5}
\end{eqnarray}
The ratio of the average momentum fractions $\bar y_1$ and $\bar y_2$ of the $u$ and $s$ type valons, respectively, should be roughly the ratio of their masses $m_U/m_S\approx 2/3$. That puts a constraint on the parameters $a$ and $b$, which is \ $b=(3a+1)/2$. Phenomenological analysis of the process $K^++p\to\pi^++X$ in the  valon model then results in $a=1$ and $b=2$ \cite{16}. For the hyperon \la\ the wave function can be written in the form \cite{6}
\begin{eqnarray}
G_\Lambda(y_1,y_2,y_3)={1 \over B(\alpha+1,\alpha+\beta+2)B(\alpha+1,\beta+1)}\ (y_1y_2)^\alpha\,y_3^\beta\ \delta(y_1+y_2+y_3-1)\ , \label{6}
\end{eqnarray}
where $y_1$ and $y_2$ refer to the non-strange valon, and $y_3$ to the strange valon. The ratio of the average momentum fractions $\bar y_1/\bar y_3$ is $(\alpha+1)/(\beta+1)$, which yields a similar constraint: $\beta=(3\alpha+1)/2.$  We suppose that their values are roughly the same as for proton \cite{15}, so they shall be set at $\alpha=1$ and $\beta=2$. 

For  \ph\ and \om\  only one type of valon is involved, so their valon \dis s have the simple symmetric forms
\begin{eqnarray}
G_{\phi}(y_1, y_2)&=&[B(c+1, c+1)]^{-1} (y_1y_2)^c \delta(y_1+y_2-1),  \label{7}  \\
G_\Omega(y_1,y_2,y_3)&=&[B(\gamma+1,\gamma+1)B(\gamma+1,2\gamma+2)]^{-1}(y_1y_2y_3)^\gamma\delta(y_1+y_2+y_3-1).  \label{8}
\end{eqnarray}
 Since the masses of those two particles are nearly at the thresholds of the sums of the constituent quark masses, loose binding means narrow wave functions in the momentum space. 
 We shall consider two cases: (a) $c=\gamma=5$, and (b) $c=\gamma=4.$ For case (a) or larger $c$ and $\gamma$, the \dis s in Eqs.\ (\ref{7}) and (\ref{8}) are narrow enough so that they are not too far from the   $\delta$-function approximations, i.e.,
\begin{eqnarray}
G_\phi(y_1,y_2)&=&\delta(y_1-1/2)\,\delta(y_2-1/2) ,   \label{7a} \\
G_\Omega(y_1,y_2,y_3)&=&\delta(y_1-1/3)\,\delta(y_2-1/3)\,\delta(y_3-1/3) .   \label{8a}
\end{eqnarray}

To calculate the \pt\ \dis s, we start with the production of $K_s^0$, for which we consider the $d\bar s$ and $s\bar d$ contributions to $F_{q\bar q'}$ and write
 \begin{eqnarray}
F_{d\bar s,s\bar d}={\cal T}{\cal T}_s+{\cal T}_s({\cal S}_d+{\cal S}_{\bar d})/2+{\cal T}{\cal S}_s+\{{\cal S}{\cal S}_s\}\ ,  \label{9}
\end{eqnarray}
where a sum over all relevant quark species is implied when not indicated explicitly.
The thermal and shower parton distributions of the light quarks are, respectively, \cite{5}
\begin{eqnarray}
{\cal T}(p_1)=p_1{dN_q^{\rm th}\over dp_1} = C\, p_1\exp (-p_1/T)\ , \label{10}\\
{\cal S}(p_2) = \xi \sum_i \int^{\infty}_{k_0}dk\, k\,
f_i(k)\, S_i (p_2/k)\ . \label{11} 
\end{eqnarray}
The parameters $C$ and $T$ have been determined in \cite{5}: 
\begin{eqnarray}
C=23.2\ {\rm GeV}^{-1} \ ,  \quad T=0.317\  {\rm GeV} \ .   \label{12}
\end{eqnarray}
$S_i$ is the SPD for a light quark in a shower initiated by a hard parton $i$, $f_i(k)$ is the transverse-momentum ($k$) \dis\ of hard parton $i$ at midrapidity in central Au+Au collisions, and $\xi$ is the average fraction of hard partons that can emerge from the dense medium to hadronize. As in \cite{5}, $f_i(k)$ is taken from Ref.\ \cite{17}, $k_0$ is set at 3 GeV/c, and $\xi$ is found to be 0.07. For the strange quarks the corresponding thermal and shower parton \dis s are 
\begin{eqnarray}
 {\cal T}_s(p_1) = p_1{dN_s^{\rm th}\over dp_1} = C_s\, p_1\exp (-p_1/T_s), \label{13}\\
 {\cal S}_s(p_2) = \xi \sum_i \int^{\infty}_{k_0}dk\, k\,
f_i(k)\, S^s_i (p_2/k)  ,
\label{14}
\end{eqnarray}
where $C_s$ and $T_s$ are now  new parameters. The SPDs $S_i^s$ are known from \cite{9} and cannot be altered. 
Indeed, $S_i$ and $S_i^s$ are so well determined from the meson fragmentation functions (FFs) that we have recently been able to calculate the baryon FFs from them in agreement with data \cite{18}, thus solidifying the validity of our approach to fragmentation in the framework of recombination.
The thermal $s$ quark distribution in Eq.\ (\ref{13}) is determined by  fitting the low-\pt\ data on $K_s^0$ production. Since $K^*$ decays totally into $\pi K$,  we have $g_K=1$.  

The last term of Eq.\ (\ref{9}) symbolizes
\begin{eqnarray}
\{ {\cal SS}_s\} = \xi \sum_i \int^{\infty}_{k_0}dk\, k\,
f_i(k)\left[\left\{ S_i^d\left({p_1\over k}\right),S^{\bar s}_i \left({p_2\over k-p_1}\right)\right\}+\left\{S_i^{\bar d}\left({p_1\over k}\right),S^{s}_i \left({p_2\over k-p_1}\right)\right\}\right]/2,
\label{15}
\end{eqnarray}
where $\{\cdots, \cdots\}$ on the RHS represents symmetrization of $p_1$ and $p_2$. 
The two terms in the square brackets are not the same because of the valence quark contribution to the $d$-initiated jet.
Despite the appearance of the product form on the LHS, there is only one integral over $k$ because both shower partons belong to the same jet inititated by one hard parton $i$. Since we sum $i$ over $u, d, s, \bar u, \bar d, \bar s$ and $g$, Eq.\ (\ref{15}) contains 28 terms. However, when they are  combined with the RF, they become the FF $D_i^{K^0_s}(p/k)$ \cite{5}.

For \la\ production we need
\begin{eqnarray}
F_{uds}&=&{\cal T}{\cal T}{\cal T}_s+{\cal T}{\cal T}_s{\cal S}+{\cal T}{\cal T}{\cal S}_s+{\cal T}_s\{{\cal S}{\cal S}\}+{\cal T}\{{\cal S}{\cal S}_s\}+\{{\cal S}{\cal S}{\cal S}_s\} \ ,  \label{16}
\end{eqnarray}
where the last term involves the symmetrization of all three shower partons in the same jet. Clearly, there are now many more terms for $F_{uds}$. We shall find that up to $p_T\approx 4$ GeV/c, the relative importance of each of the six terms in Eq.\ (\ref{16}) diminishes in the order they appear on the RHS. Nevertheless, they all have to be calculated in order to determine the emergence of certain terms, as \pt\ is increased. For the RF of \la\ the statistical factor $g_\Lambda$ is 1/4.
 With the parameter $\alpha$ in the \la\ wave function expressed in Eq.\ (\ref{6}) set at 1 to simulate the proton, there are no adjustable parameters in this \la\ production problem.

For \ph\ and \om\ the \pt\ \dis s can be written out explicitly if the naively simple forms in Eqs.\ (\ref{7a}) and (\ref{8a}) for their RFs are used. They are 
  \begin{eqnarray}
{dN_\phi\over pdp}&=&{g_\phi \over pp_0}\,F_{s\bar s}(p/2, p/2) ,   \label{17}\\
{dN_\Omega\over pdp}&=&{g_\Omega\over pp_0}\,F_{sss}(p/3, p/3, p/3) ,  \label{18}
\end{eqnarray}
where
 \begin{eqnarray}
F_{s\bar s}&=&{\cal T}_s{\cal T}_s+{\cal T}_s{\cal S}_s+\{{\cal S}_s{\cal S}_s\},   \label{19}\\
F_{sss}&=&{\cal T}_s{\cal T}_s{\cal T}_s+{\cal T}_s{\cal T}_s{\cal S}_s+{\cal T}_s\{ {\cal S}_s{\cal S}_s\}+\{{\cal S}_s{\cal S}_s{\cal S}_s\} \ .   \label{20}
\end{eqnarray}
In our calculation we shall, however, use the RFs given in Eqs.\ (\ref{7}) and (\ref{8}) and consider the two cases given below those equations.
We allow the normalization factors $g_\phi$ and $g_\Omega$  to be adjustable in order to account for the medium effect on the recombination of $s\bar s$ and $sss$. Because of the limited supply of $s$ quarks and the abundance of the light quarks in the vicinity, it is expected that the formation of  the  multi-strange particles at higher masses would be suppressed. Without a principle from outside the recombination model, there is no way to determine the suppression factors. In the following we  vary $g_\phi$ and $g_\Omega$ as free parameters to fit the normalizations of the corresponding spectra. 
Current data on the spectra of strange particles  extend to  \pt\ as high as 6 GeV/c  \cite{18}-\cite{25}.

 \section{Results}
 
 We first show the result of our calculation for $K_s^0$ production. In Fig.\ 1 is shown the contributions from the three components indicated in the figure. The thermal-shower component includes both the ${\cal T}_s{\cal S}$ and ${\cal TS}_s$ subcomponents, although the latter subcomponent is much smaller because the shower $s$ quark is suppressed. The contribution from ${\cal T}_s{\cal S}$ is more  important because thermal $s$ quarks are not negligible, as evidenced by the ${\cal TT}_s$ component that is dominant for $p_T<3$ GeV/c. The $\{{\cal SS}_s\}$ component does not become important until $p_T>8$ GeV/c.
 
 In fitting the low-$p_T$ region for $p_T<2$ GeV/c we use 
  \begin{eqnarray}
C_s=13.9\  {\rm GeV}^{-1}\ ,  \qquad T_s= 0.33\  {\rm GeV} \ .    \label{21}
\end{eqnarray}
 This value of  $T_s$  is slightly higher than $T=0.317$ GeV for light quarks.  Since the low \pt\ phenomenology includes the flow effect, which we do not separately put in, that small difference in the inverse slopes is sensible in view of the mass difference involved.
Note that it is because of the importance of the contribution of the thermal-shower recombination for $p_T>3$ GeV/c that the \pt\ \dis\ bends up from the exponential behavior to agree with the data  \cite{23}. 

Having determined $C_s$ and $T_s$, we can now proceed to the calculation of \la\ production without adjusting any parameters. There are, however, two cases to consider that are related to the valence quarks in the showers. The question is whether the leading parton is likely to form a leading meson in a jet or a baryon with higher mass, remembering that there is only one valence quark in a quark-initiated shower. We calculate the \pt\ \dis s of \la\ for the two cases, where (a) only the sea quarks in the showers are used for the formation of \la, and (b) all contributions including all valence quarks are included. The results  are shown in Fig.\ 2(a) and (b), respectively.  There is a small but perceptible difference between the two cases for $p_T>4$ GeV/c. The inclusion of valence quarks in (b) naturally has the effect of increasing the yield at high \pt.

In Fig.\ 2 we first note that the recombination of three thermal partons (TTT) is dominant up to about 4 GeV/c, where the TTS contribution becomes more important at higher \pt.  The TSS component is unimportant below 6 GeV/c, and SSS is even less important.  Fig.\ 2(a)  gives a more  satisfactory reproduction of the data \cite{23} than Fig.\ 2(b). It implies that the valence quarks that have larger momentum fractions in a jet are mainly in the mode of formation of mesons instead of hyperons. Gluon jets make the dominant contribution to the shower partons that form the \la.

With the calculated \pt\ \dis s of $K_s^0$ and \la\ at hand, we can take their ratio and compare it with the data on $\Lambda/K_s^0$ \cite{22,23}. Since the latter is shown in linear scale, it is far more sensitive to the details of the \pt\ \dis s than the spectra themselves plotted in log scale. 
What is shown by the solid line in Fig.\ 3 is the result on the $\Lambda/K_s^0$ ratio using the \pt\ \dis s shown in Figs.\ 1 and  2(a).
Our calculated result in Fig.\ 3 does not reproduce  the data at large  $p_T$, although they agree well for $p_T<2$ GeV/c. The solid line has the right structure of a peak reaching almost as high as 2. The disagreement with data is an amplification of the small discrepancies between theory and experiment in Figs.\ 1 and 2(a), where our calculated \dis\ for $K_s^0$ in the region $4<p_T<6$ GeV/c is slightly lower than the data, but for \la\ it is slightly higher than the data. That intermediate \pt\  region is where the shower contributions become important. Evidently, our model for the production of such strange particles is not good enough to yield better than 30\% accuracy. 
When the shower parton contribution is not strong, it is necessary to rely on the accuracy of the thermal parton distribution extrapolated to very high \pt, higher than can perhaps be justified with the simple exponential formula used. Moreover, it  is likely that hadronization into different strange hadronic channels may not be independent, so the partition problem in the strange sector may have to be taken into account.

For the production of \ph\ and \om\ we have done the calculation for both cases (a) $c=\gamma=5$ and (b) $c=\gamma=4$ for the valon \dis s given in Eqs.\ (\ref{7}) and (\ref{8}) that are used in the RFs in Eqs.\ (\ref{3}) and (\ref{4}). For clarity we show only the results for case (a) in Figs.\ 4 and 5, since the other case will be shown to be ruled out by the $\Omega/\phi$ ratio.  If we ignore all but the first terms on the RHS of Eqs.\ (\ref{19}) and (\ref{20}), we would have only the thermal contributions, which from (\ref{17}), (\ref{18}) and (\ref{13}) give\footnote{These simple formulas are the result of using the $\delta$-function approximations of the wave functions in Eqs.\ (\ref{7a}) and (\ref{8a}) for the recombination functions.} 
 \begin{eqnarray}
{dN_\phi\over pdp}=g_\phi C_s^2 {p\over 4p_0} e^{-p/T_s} \ ,    \label{22}\\
{dN_\Omega\over pdp}=g_\Omega C_s^3 {p^2\over 27p_0} e^{-p/T_s} \ ,  \label{23}
\end{eqnarray}
where $C_s$ and $T_s$ are as given in  Eq.\ (\ref{21}).  The results from using $c=\gamma=5$ in Eqs.\ (\ref{7}) and (\ref{8}) are similar, but cannot be expressed in simple analytic form as in Eqs.\ (\ref{22}) and (\ref{23}); they are shown by the dashed lines in Figs.\ 4 and 5. Being from the thermal contributions only, they are nearly exponential for $p_T>2$ GeV/c.  The other terms in Eqs.\ (\ref{19}) and (\ref{20}) involving shower partons give rise to contributions shown by the other lines in those figures, which are lower.  Their sums represented by the  solid lines differ only slightly from the dashed lines. It is clear then from Figs.\ 4 and 5 that the \pt\ \dis s of both \ph\ and  \om\ are dominated by the recombination of thermal partons $({\cal T}_s)$ in the regions where  data are available \cite{21,24,25}. The contribution of thermal-shower recombination $({\cal T}_s{\cal S}_s)$ to \ph\ production does not exceed that from ${\cal T}_s{\cal T}_s$ until $p_T>6$ GeV/c. For \om\ the cross-over is above 8 GeV/c, not shown in Fig. 5. The production  of shower $s$ quark is suppressed, but it is enough to make a few percentage contribution at the high end of \pt\ where data exist.\footnote{This result differs from the earlier one in \cite{26}, where a number of approximations were made, and a smaller values of $\xi$ was used by mistake.}
The overall normalizations of the \dis s are fitted by adjusting the values of $g_\phi$ and $g_\Omega$  to be
\begin{eqnarray}
g_\phi=0.34\ , \quad  g_\Omega=0.012\ .    \label{24}
\end{eqnarray}
They represent the degree of suppression of the rate of recombination of 
an $s$ quark  with an $\bar s$ (to form \ph) and two other $s$ quarks (to form \om) in the environment of light quarks that present competing channels for the formation of kaons and hyperons. The phenomenon is similar to that in the \la\ production where the leading $s$ quark prefers to form $K$ rather than \la, as shown in the two parts of Fig.\ 2.

To calculate the ratio $R_{\Omega/\phi}$ we can start with the simplest by using Eqs.\ (\ref{22}) and (\ref{23}) and get 
\begin{eqnarray}
R_{\Omega/\phi}^{\rm th}(p)={4g_\Omega C_s\over 27 g_\phi} p ,  \label{27}
\end{eqnarray}
which is linearly rising. However, that is for thermal-parton recombination with $\delta$-function RFs. For case (a) RFs with $c=\gamma=5$, the result for thermal contribution only is shown by the dashed line in Fig.\ 6, which is the ratio of the dashed lines in Figs.\ 4 and 5, and  is approximately linearly rising. When all terms are included, the ratio of the solid lines in Figs.\ 4 and 5 results in the solid line in Fig.\ 6. There is a definitive bending down of the ratio for $p_T>5$ GeV/c, although the solid lines in Fig.\ 4 and 5 deviate only slightly from the dashed lines in the log scale. The linear scale in which the ratio is shown magnifies the deviation arising from the contributions of the shower partons. The solid line exhibits the trend of the data that departs from linearity at $p_T\approx 4$ GeV/c \cite{25}; however, our model calculation gives a peak at $p_T\approx 5$ GeV/c which is slightly higher than the data, and with the maximum at $R_{\Omega/\phi}\approx 0.25$. The decrease of \op\ at high \pt\ is due to the increase of the \ph\ spectrum that is lifted up from the exponential behavior by the contribution from thermal-shower recombination. That upward bending of the \ph\ spectrum occurs   earlier in \pt\ than that of  \om, with the consequence that \op\ decreases for $p_T\ge 5$ GeV/c. In Fig.\ 6 we also show in dashed-dotted line the $\Omega/\phi$ ratio for $c=\gamma=4$ in the wave function of \ph\ and \om\ [see Eqs.\ (\ref{7}) and (\ref{8})], corresponding to broader widths in the momentum-fraction space. The result is far from the data and should be discarded from further consideration. It does confirm our earlier statement that narrow width in the momentum space is expected due to the weak binding of constituent quarks in \ph\ and \om.

 \section {Discussion}

In the approach to the hadronization problem that we have adopted there are certain properties that we cannot change, while there are others that we can determine only by fitting the data. The former involve the hard partons and their shower parton distributions. Of course, they can be improved by going to higher order calculations, which we rule out for the present consideration. The latter involve the soft thermal partons, for which several parameters have been used for non-strange and strange quarks. Since our aim has never been to model the soft component, it does not matter how many parameters have been used to reproduce the low \pt\ behavior. Our aim has been to understand strange particle production at intermediate and high \pt, and
we have found some valuable information to provide significant insight into the strange sector.  The recombination model has been effective in illuminating the physics involved.

From $K_s^0$ and \la\ spectra we have seen how the deviation from the exponential behavior that describes the thermal components can be understood as being due to the emergence
 of the shower contribution arising from jets. It happens at around $p_T\approx 4$ GeV/c. Those are  strange hadrons containing light quarks, which are easier to produce at high \pt\ than the strange quarks that are suppressed. The thermal source provides the $s$ quarks for recombination. The $\Lambda/K$ ratio has a maximum at $p_T\approx 4$ GeV/c, a bit higher than the location of the peak in the data \cite{23}. The cause of the decrease of the ratio is the relative ease of $K$ over $\Lambda$ in acquiring a boost in yield by thermal-shower recombination at high \pt. Quantitatively, we do not expect better than 30\% accuracy in the spectra themselves, since lowest order pQCD was used in the calculation of hard scattering, and the virtuality of the shower partons is not changed according to the \pt\ of the fragmenting hard partons. Despite that limitation the fact that the spectra can be reproduced over such a wide range of yield for \pt\ up to 6 GeV/c (a span of 5 orders of magnitude) is sufficient to be regarded as support for the validity of the recombination model used.
 
In the case of \ph\ and \om\ the absence of light quarks in them deprives them of the light shower partons at high \pt, so their productions are dominated by the thermal strange-quark recombination. The contribution from hard parton scattering is negligible unless \pt\ is large. The small upward bend of the \ph\ and \om\ spectra above straight-line behaviors in the log plots (i.e., exponential) results in the $\Omega/\phi$ ratio to bend down above $p_T\approx 5$ GeV/c. \om\ is harder to be lifted up by shower $s$ quark than \ph\ because there are two thermal $s$ quarks instead of one, given that shower $s$ partons are suppressed in the first place.

The significance of the \op\ ratio in Fig.\ 6 is not so much the bending down of the solid line at high \pt\ as the rising of the dashed line at intermediate \pt\ in agreement with the data. That rising behavior can be seen clearly in the simple analytic expressions in Eqs.\ (\ref{22}) and (\ref{23}), from which one gets the ratio to be proportional to $p$ (which is abbreviation for  \pt\ in Fig.\ 6). Although those expressions are simplified approximations, they exhibit two features: (a) the extra factor of $p$ in Eq.\ (\ref{23}) compared to (\ref{22}) is due to the difference in the RFs of the baryon versus meson shown in Eqs.\ (\ref{4}) and (\ref{3}), respectively, and (b) the more important aspect of those two expressions is the common exponential factors that are cancelled in the ratio --- those factors are dynamical in origin, while the prefactors are kinematical.  The fact that the dashed line in Fig.\ 6 well describes the data up to $p_T\approx 4$ GeV/c implies that both \ph\ and \om\ have the same dynamical origin of thermal partons, which are assumed to have exponential behavior given in Eq.\ (\ref{13}). The surprise is the validity of that exponential form out to $p\approx 2$ GeV/c for \ph\ observed at $p_T\approx 4$ GeV/c, remembering that $T_s$ is only 0.33 GeV. Thus we have a very robust system of thermal partons consisting of not only the light quarks but also the strange quarks. That is the main characteristic of quark-gluon plasma \cite{1,2}. It takes the production of \ph\ and \om\ to reveal cleanly this thermal source because their exponential behavior at intermediate \pt\ is not covered up by the contamination of shower partons from hard scattering, as is the case with other mesons and baryons. In that sense \ph\ and \om\ serve as windows through which one can observe directly the properties of the partonic source.

The discussion above concerns the shapes of the \pt\ \dis s of \ph\ and \om. The normalizations have been adjusted to fit the data by giving the parameters $g_\phi$ and $g_\Omega$ the values shown in Eq.\ (\ref{24}). The fact that they are less than 1 implies suppression. We interpret that as being due to the competition for $s$ quarks in the environment of light quarks during hadronization. Evidently, it is less likely for an $s$ to find an $\bar s$ to form \ph\ than a $\bar u$ or $\bar d$ to form $K$; it is even more difficult to find two other $s$ quarks to form \om. To study the constraints arising from competing channels is outside of the scope of this paper, since our emphasis has been on the \pt\ dependence of the strange hadron production. The values of $g_\phi$ and $g_\Omega$ can, however, be useful at a later point when the competiton problem is considered.

The phenomenological fact that the \pt\ \dis\ of \om\ is essentially exponential up to $p_T\approx 6$ GeV/c leads one to conclude, even without the benefit of our model calculation, that \om\ is produced from the thermal source. It is then natural to predict that any particle produced in association with \om\ must also be from the same thermal source, and hence indistinguishable from the background. Such a prediction was, in fact, made in the unpublished version of this work \cite{26}, which stimulated an experimental search for the partners of \om\ above background. The result of that study was reported at Quark Matter 2006, indicating that unidentified charged particles were found on the near side of the $\Delta\phi$ \dis\ in association with \om\ as trigger for $2.5<p_T^{\rm trig}<4.5$ GeV/c and $1.5<p_T^{\rm assoc}<p_T^{\rm trig}$ \cite{27}.
Those are rather low values of $p_T^{\rm trig}$ and $p_T^{\rm assoc}$,  lower than what can unambiguously give rise to the usual jet structure. For $p_T^{\rm trig}<4.5$ GeV/c we see in Fig.\ 5 that the contribution from thermal-thermal-shower recombination is minuscule compared to that of thermal partons only. Thus there are no apparent jets to account for the associated particles. This dilemma, called the \om\ puzzle, was addressed at Quark Matter 2006 and resolved conceptually by the notion of phantom jet \cite{28}. Without entering into the details of the possible resolution, we mention here only the essence of the idea. It is conjectured that both the \om\ and its associated particles are produced from the  ridge induced by jets, the nature of which has been studied for other triggers \cite{27}.   The ridge is an enhanced thermal source, and is itself not jet-like, but without a jet (created by gluon or light quark) there would not be any ridge \cite{29}. A ridge without peak may therefore be regarded as the shadow of a phantom jet. If the  associated particles are in the ridge, then they can be observed above the background.
Quantitative implementation of this idea remains to be carried out, but probably not until more data become available on ridge characteristics. At present, without $\Delta\eta$ distribution there is no identification of a ridge due to the low statistics in \om\ production.

In conclusion, we remark  that intermediate \pt\ physics in the strange sector at RHIC has been substantially illuminated by this study where hadron production up to $p_T\sim 6$ GeV/c can be understood in terms of recombination. Fragmentation is ruled out. Although the calculated results on $\Lambda/K$ and \op\ ratios do not agree perfectly with the data on the high end of the \pt\ range, they represent only less than 30\% discrepencies in the spectra themselves that vary over 5 orders of magnitude. The most significant revelation is that a robust thermal parton source is created at RHIC before hadronization.

\section*{Acknowledgment} 
 We thank Nu Xu for valuable discussions and to Peter Jacob for helpful communication in the early stage of this work. We are also grateful to Betty Abelev, Jana Bielcikova, Sarah Blyth and Rene Bellwied for keeping us informed on the progress of their data analyses and for answering many questions. This work was supported, in part,  by the
U.\ S.\ Department of Energy under Grant No. DE-FG02-92ER40972   and by National Natural Science Foundation of China under Grant No. 10475032.

\newpage
\centerline{\large{\bf Figure Captions}}
\vskip0.5cm
\begin{description}
\item 
Fig.\ 1. Transverse momentum distribution of $K_s^0$ in central Au+Au
collisions. Data are from \cite{23}. The solid line is the sum of
the three contributions: $\cal TT$ (dashed
line), $\cal TS$ (dashed-dot line),  $\cal SS$ (dotted line).

\item
Fig.\ 2. Transverse momentum distribution of \la\ in central Au+Au
collisions. Data are from \cite{23}. The heavy solid line is the sum of
the four contributions: $\cal TTT$ (dashed line), $\cal TTS$ (dashed-dotted
line), $\cal TSS$ (dotted line),  $\cal SSS$ (light solid line). (a) no valence quarks are included; (b) valence quarks are included.

\item
Fig.\ 3. The ratio of \la\ to $K_s^0$. Data are from \cite{22,23}. The solid line is from Fig.\ 1 and Fig.\ 2 (a).

\item 
Fig.\ 4. Transverse momentum distribution of \ph\ in central Au+Au
collisions. Data are from \cite{19,20}. The solid line is the sum of
the three contributions: $\cal TT$ (dashed
line), $\cal TS$ (dashed-dotted line),  $\cal SS$ (dotted line).  

 \item 
 Fig.\ 5. Transverse momentum distribution of \om\ in central Au+Au
collisions. Data are from \cite{21,24,25}. The heavy solid line is the sum of
the four contributions: $\cal TTT$ (dashed
line), $\cal TTS$ (dashed-dotted line), $\cal TSS$ (light solid line),  $\cal SSS$ (dotted line).  

\item
Fig.\ 6. The ratio of \om\ to \ph. Data are from \cite{24,25}. The dashed line is for $c=\gamma=5$ in the RF and for thermal parton recombination only. The solid line is for $c=\gamma=5$ and includes all components of recombination. The dashed-dotted line is for $c=\gamma=4$ with all terms included.

\end{description}
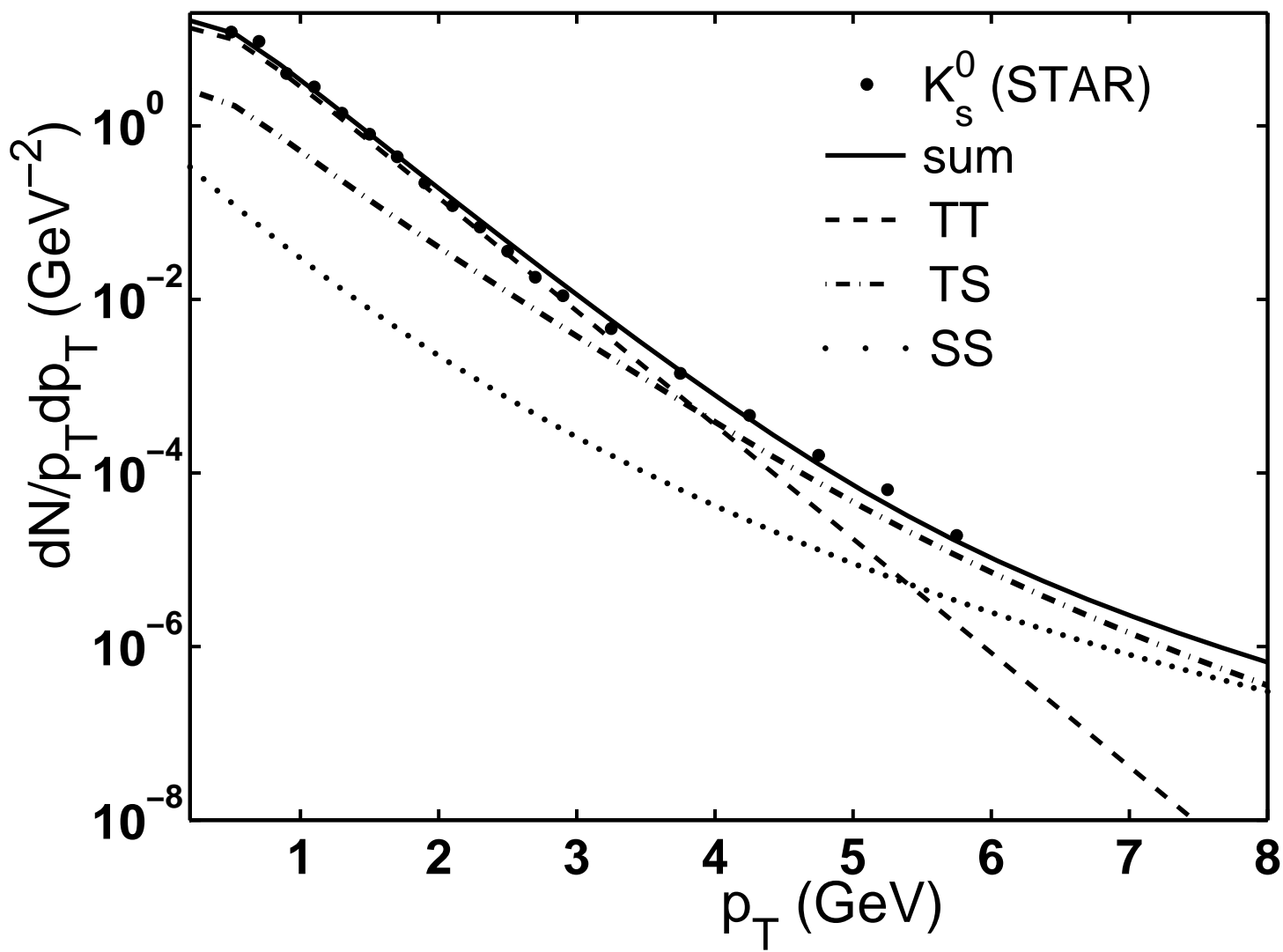
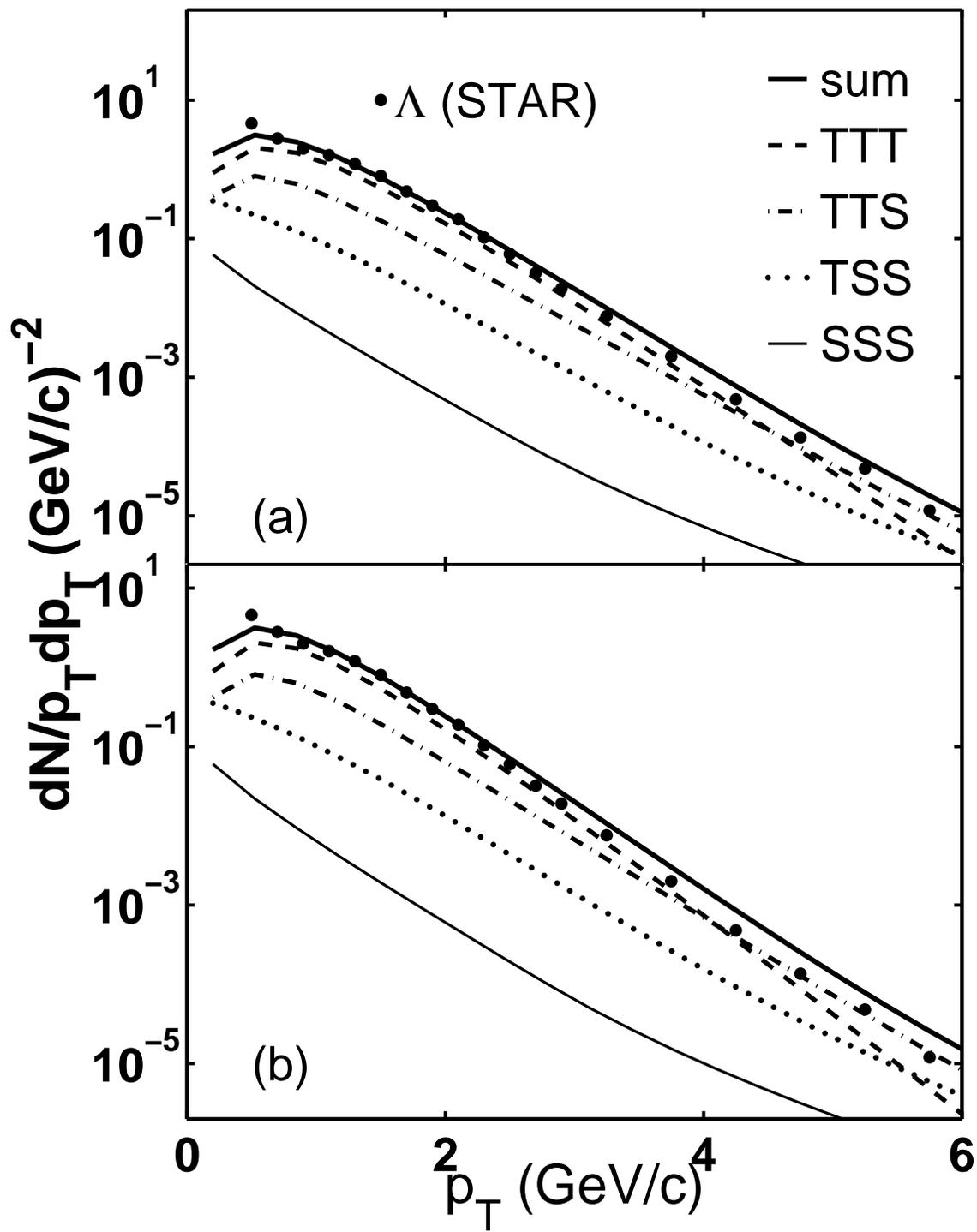
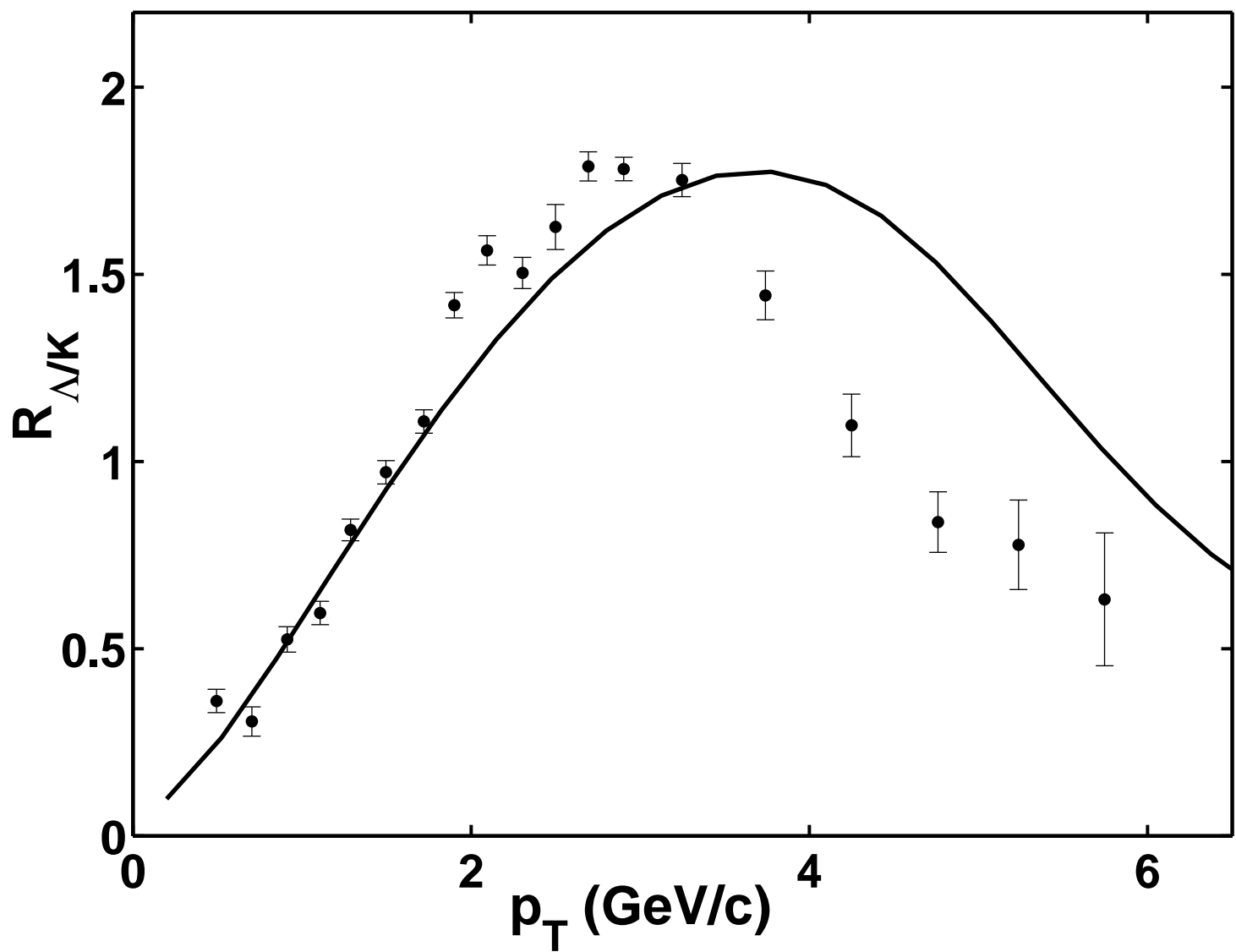
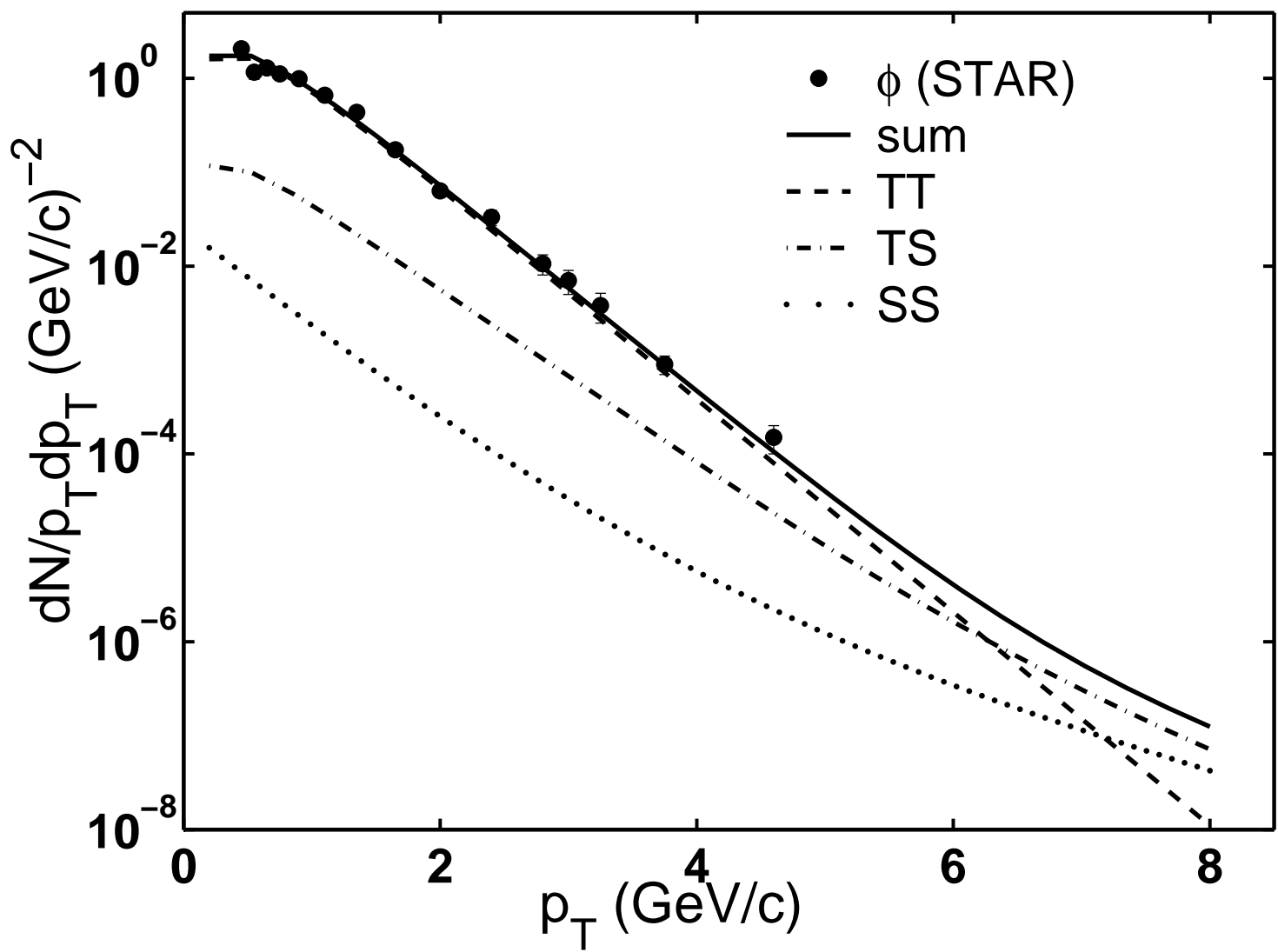
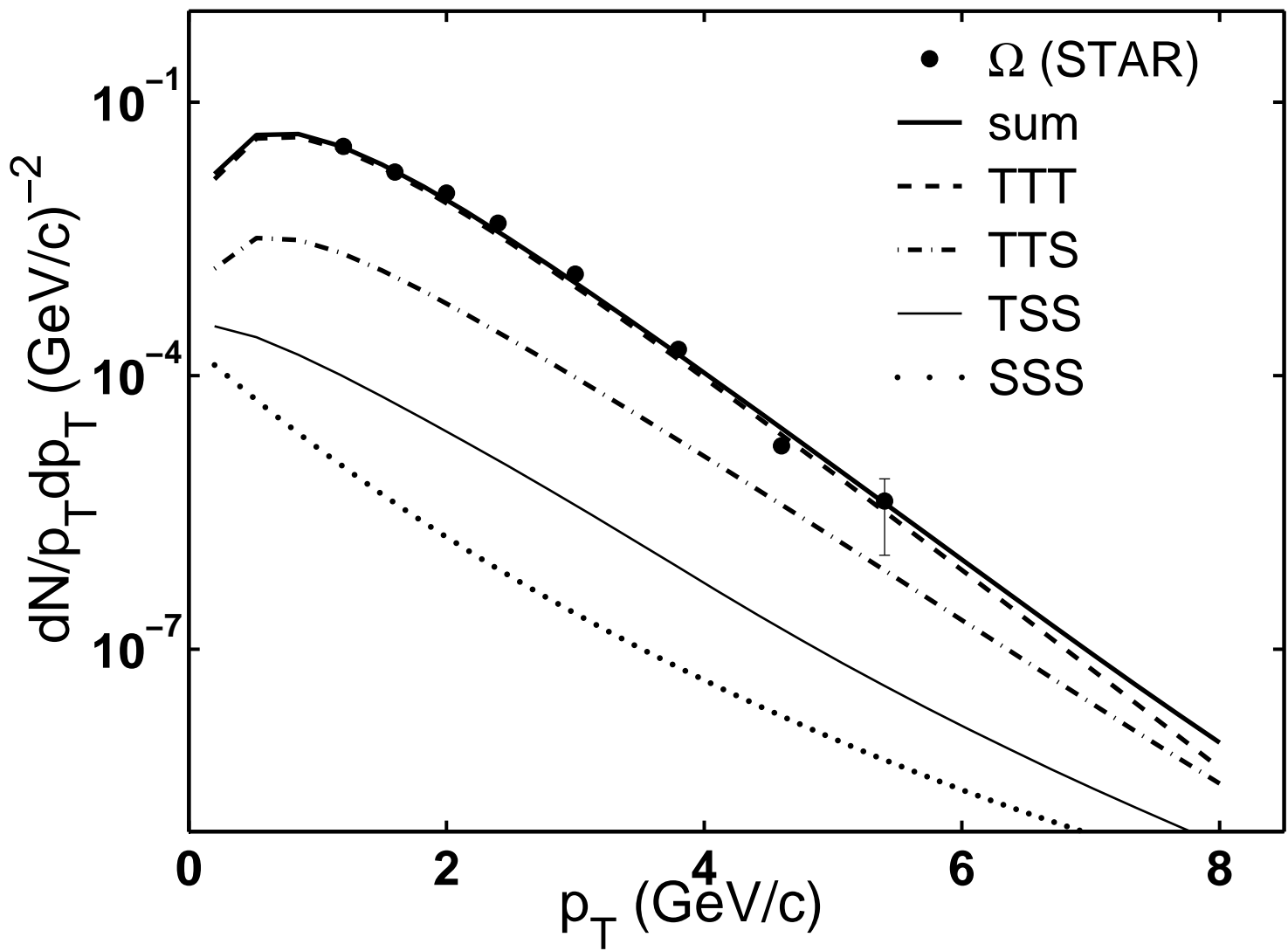
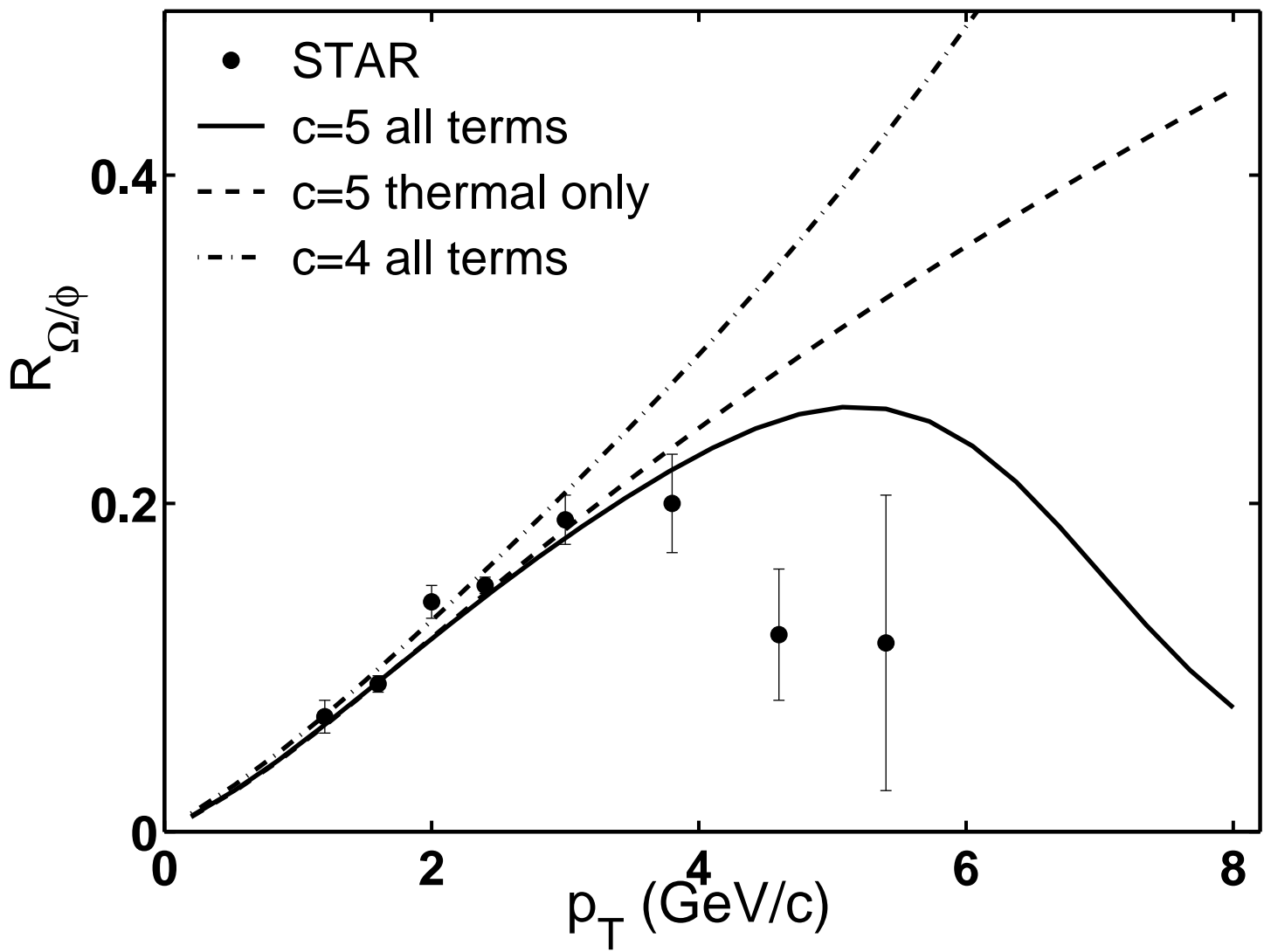
\end{document}